\documentstyle[twocolumn,epsfig,prl,aps,amsmath,amssymb,float]{revtex}
\begin{document}
\newfloat{figure}{ht}{aux}
%\preprint{IUCM96-019}
\draft
\twocolumn[\hsize\textwidth\columnwidth\hsize\csname
@twocolumnfalse\endcsname
\title{Logrithmic corrections to the RG flow for the two-dimensional
bond disordered Ising model}
\author{E.~S.~S\o rensen}
\address{Laboratoire de Physique Quantique \& 
UMR CNRS 5626, Universit\'e Paul Sabatier, 31062 Toulouse, France}  
\date{\today}
\maketitle
\begin{abstract}
Using the mapping of the partition function of the two-dimensional
Ising model onto a pfaffian we evaluate
the domain wall free energy difference for the pure and disordered Ising
model close to the pure fixed point. Using this method very large
lattices can be studied exactly and we confirm that disorder even
including
frustrating interactions indeed are irrelevant close to the pure fixed
point. The finite-size renormalization group flow shows a power-law
behavior modified by a logarithmic term that dominates for small lattice
sizes.
\end{abstract}
\vskip2pc]
%============================================================================
% BODY OF PAPER

\section{Introduction}
The influence of disorder on the pure two-dimensional Ising model has
been a problem of long standing interest. The works of
Ludwig~\cite{ludwig}, Shankar~\cite{shankar} and Dotsenko and
Dotsenko~\cite{DD} have shown that logarithmic corrections to
the specific heat arises once disorder is present but the otherwise
the critical exponents remain unchanged from the pure case. In a
certain sense one might say that disorder is marginally irrelevant for the pure
Ising model. From the Harris criterion~\cite{harris} we know that
disorder is relevant if the correlation length exponent, $\nu > 2/d$.
However, for the two-dimensional Ising model $\nu=1$ and the Harris
criterion gives no definite answer to relevance or irrelevance of
disorder which in this case is a marginal perturbation. Most of the
analytical treatments of this problem considers only non-frustrating
disorder and one might ask the question if frustration could
have a different influence on the system than non-frustrating disorder.

In this paper we revisit this problem and by mapping the partition
function onto a pfaffian we are able to obtain large scale numerically
exact results independent of the disorder distribution for various
different quantities close the pure fixed point. To our knowledge
pfaffian techniques have so far not been applied to fully disordered
systems. Compared to more standard transfer matric techniques
this method presents significant advantages notably
in the system sizes it is possible to treat. In some cases it is
possible to obtain exact results for system sizes comparable to what
has been treated with Monte Carlo methods.
Using finite-size
scaling ideas we show that disorder indeed is irrelevant at the
pure fixed point even including frustration. However, the finite-size
renormalization group flow has a novel logarithmic behavior recently 
predicted to be rather universal for many models by Cardy~\cite{cardy}.
For small system sizes disorder appears to be relevant and only
for very large system sizes does it become apparent that the flow
is towards the pure fixed point.

\section{method}
Our approach is inspired by the domain wall renormalization group (DWRG)
developed by McMillan~\cite{mcmillan} and a recent modification of
this, used to determine the fixed point structure of the two-dimensional
disordered Potts model~\cite{us}. In this approach the stiffness of the
system is calculated and a finite size scaling ansatz is used to
determine the flow. We briefly recapitulate the central results. 
In the absence of disorder
it is known from hyperscaling that the singular
part of the free energy density 
scales as the inverse of a correlation volume
\begin{equation}
\frac{f_s}{k_bT_c} = \frac{C}{\xi^d}.
\label{eq:hscaling}
\end{equation}
Exploting the point of view of the DWRG we assume that the critical part
of the free energy, $f_s$, comes
from the difference in free energy between two configurations with
different boundary conditions $f_s=\Delta f= f_{\rm periodic}-f_{\rm
antiperiodic}$. Using standard finite-size scaling arguments~\cite{us}
we find that the total free energy difference $\Delta F$ scales as
\begin{equation}
\frac{\Delta { F}}{k_bT} = A g(\delta L^{1/\nu}),
\label{eq:scalepure}
\end{equation}
with $g$ a universal function and $\delta=|T-T_c|$.
At $T_c$ the free enrgy difference is thus a universal amplitude.

$A g(\delta L^{1/\nu})$ can be regarded as the stiffness of the
system, $\rho(L,T)$. In an ordered phase we would expext $\rho$
to diverge with the system size, whereas in a disordered phase 
$\rho$ should scale to zero with the system size.
Deep in the ordered phase it is known that the leading correction
to $\Delta F$ diverges linearly with the system size $L$~\cite{MW}.
\begin{equation}
|\Delta F/T|\simeq L\left|\ln\left(\frac{1}{|z|}\frac{1-|z|}{1+|z|}\right)\right|, \
\ z=\tanh\left(\frac{J}{T}\right).
\end{equation}
At $T=T_c$ this term is rigourously zero and the finite universal free
energy difference comes from higher order corrections.
The critical point, $T_c$, can now be located by 
standard methods, i.e. by tracing
$\Delta F$ as a function of $T$ for several different system sizes
and locating the point where the lines cross.

The Hamiltonian that we consider is the bond disordered two dimensional
Ising model:
\begin{equation}
H=-\sum_{<i,j>}
J_{ij}S_iS_j,\ \ S_i=\pm 1.
\end{equation}
To facilitate some of our calculations we restrict our treatment to
a bimodal distribution of the bonds:
\begin{equation}
{\cal P}(J) =x\delta (J - 1) + (1-x)\delta (J - 1),
\end{equation}
where $x$ is the concentration of antiferromagnetic bonds.
However, one should note that the present techniques can be used for any
distribution of disorder, continuous as well as discrete.

%%%
%% FIG 0
%%%
\begin{figure}
\begin{center}
\epsfig{file=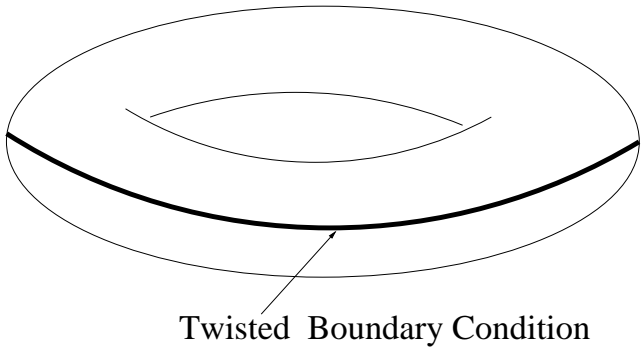,width=8.5cm}
\caption{ The antiferromagnetic seam on the torus.}
\label{fig:method}
\end{center}
\end{figure}
Since the works of Kac and Ward~\cite{KW} it has been known that the
partition function for some two-dimensional models can be expressed
as a pfaffian. A general theorem due to Kasteleyn~\cite{Kasteleyn} shows
that this is possible for all planar graphs. 
%Subsequent work by
%Little~\cite{Little} showed that this is possible for all $K_{3,3}$-free
%graphs. 
For the problem at interest here this means that a pfaffian
formulation is not possible in higher dimensions.
The Pfaffian formulation for the two-dimensional Ising model  was later
developed in detail by several authors~\cite{KW,Fisher,HG,MW}. We refer the reader to
the litterature for the derivation of this mapping.
What is very noteworthy is that the
the mapping onto a pfaffian can be exploited without problems 
even for the fully
disordered case.
For the disordered Ising model on a torus one finds a sum of four pfaffians:
\begin{eqnarray}
Z&=&2^{L^2}(\cosh(J/T))^{2L^2-N}(\cosh(-J/T))^N\nonumber\\
& &(-Pf_1+Pf_2+Pf_3+Pf_4)/2,
\end{eqnarray}
with $N$ the number of AF bonds.
In order for this expression to be useful it is necessary to be able
to evaluate the pfaffians for rather large matrices. Usually this is
done by using the well known relation for a skew-symmetric matrix $D$,
$|{\rm Pf}(D)|={\rm Det}(D)^{1/2}$. However, even though efficient
methods exists for evaluating determinants, this expression only
determines the absolute value of the pfaffian. Fortunately, recent
developments in combinatorial algorithms have shown that it is possible
to evaluate the pfaffian directly in polynomial time~\cite{Meena},
without making use of the above formula. This polynomial algorithm is
relatively slow compared to the more efficient methods for evaluating
the
determinant and for the larger systems sizes we have considered we
have calculated the determinant and obtained the sign using continuity
arguments. This can be done since the sign of the four pfaffians is
positive at sufficiently high temperatures.

\section{Results - Pure System}
As an illustration of how well this works we first consider the pure
Ising model without disorder and try to locate the known critical
point, $T_c=2/\log(1+\sqrt{2})$, using the above mentioned techniques.
Introducing a line of antiferromagnetic bonds along one line through the
system, as illustrated in Fig.~\ref{fig:method}, we can calculate the 
total free energy difference, $\Delta F$, between periodic and antiperiodic boundary
conditions. 
%%%
%% FIG 1
%%%
\begin{figure}
\begin{center}
\epsfig{file=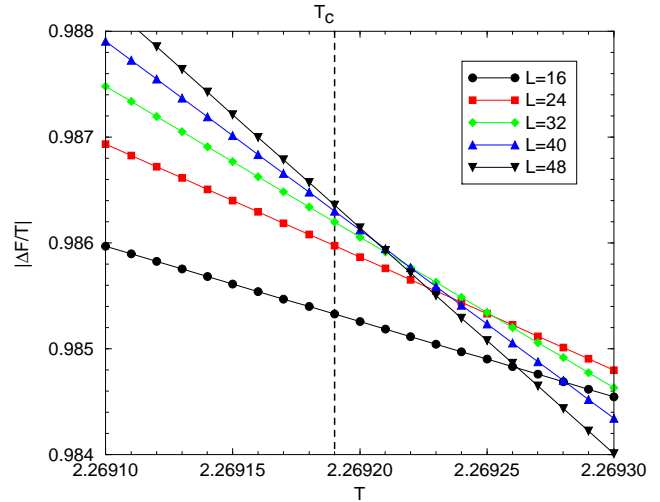,width=8.5cm}
\caption{ $|\Delta$F/T$|$ for $L=16,24,32,40,48$ as a function
of T. 
}
\label{fig:clean}
\end{center}
\end{figure}
Our results are shown in Fig.~\ref{fig:clean} for system sizes of $L=16,24,32,40,48$
as a function of T. One should note that the scale
on the x-axis which is extremely close to the critical point,
$T_c=2.269185\ldots$. On this very fine scale it is clear that
corrections to scaling are present. Only in the thermodynamic limit do
we approach the universal value of the stiffness at the critical point.
However, even neglecting these corrections, the critical point can be
located to four decimal plase. The slope of $\Delta F$ at $T_c$ should
scale as $L^{1/\nu}$ and can be used for a very precise determination of
the exponent $\nu$.
%%%
%% FIG 2
%%%
\begin{figure}
\begin{center}
\epsfig{file=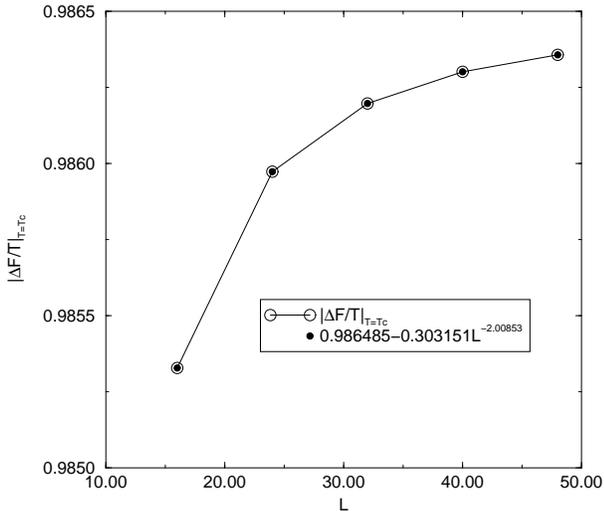,width=8cm}
\caption{$|\Delta$F/T$|$ at $T=T_c$ for $L=16,24,32,40,48$. The open
circles ($\circ$) indicate the exact numerical results and the solid
circles ($\bullet$) indicate a power-law fit to the data of the form
$0.986485-0.303151 L^{-2.00853}$.
}
\label{fig:dftc}
\end{center}
\end{figure}
Since $T_c$ in this case is known we can explicitly study the
corrections to scaling at $T_c$ by simply calculating $\Delta
F_{T=T_c}$. Our results are shown in Fig.~\ref{fig:dftc}. A simple
power law fit to the exact numerical results works extremely well and
we find that $\Delta F_{T=T_c}$ approaches a universal constant in 
the following manner:
\begin{equation}
|\Delta F/T |\simeq 0.986485-0.303151 L^{-2.00853}.
\end{equation}
From conformal invariance it is known that for the geometry of
an infinite long cylinder 
$\Delta F/T=\pi \eta$~\cite{cardy2}, where $\eta$ is the magnetic
critical exponent. Hence, for this geometry, $\Delta F$ directly
measures a {\it bulk} critical exponent.
However, for the toroidal geometry used here we are not
aware of any results for the universal number, $0.986485$, even
though many exact properties are known for the Ising model. We would
welcome any information as to where and if such a calculation has
been performed. The corrections to scaling follow a standard power-law
form with an exponent extremely close to $2$ as expected.

%%%
%% FIG 3
%%%
\begin{figure}
\begin{center}
\epsfig{file=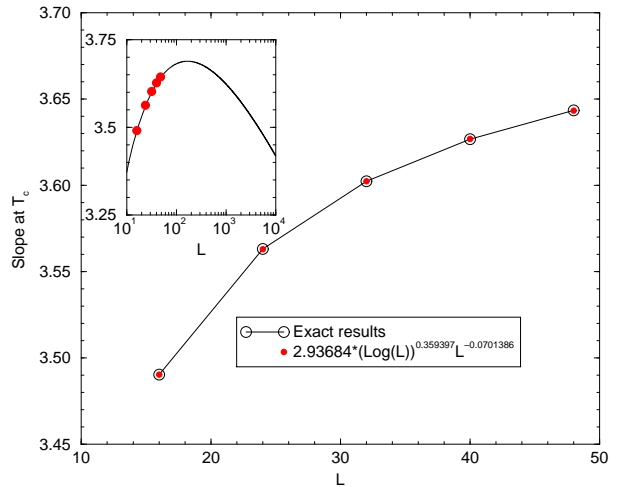,width=8cm}
\caption{The slope at $T=T_c$ for $L=16,24,32,40,48$. The open
circles ($\circ$) indicate the exact numerical results and the solid
circles ($\bullet$) indicate a power-law fit to the data of the form
$2.94\log(L)^{0.36}L^{-0.070}$.
In the inset we show the same results as well as the extrapolated 
form out to $L=10^3$.
}
\label{fig:slope}
\end{center}
\end{figure}
\section{Results - Disordered System}
We now turn to a discussion of our results in the presence of disorder.
In this case we need to consider not only $\Delta F$ but also its
standard deviation $\sigma(\Delta F)$. The relevant variable to consider
is then $r=\sigma(\Delta F)/[\Delta F]$ where $[\cdot ]$ denotes the
disorder average. We are interested in studying the behavior of this
quantity close to the pure fixed point where $\sigma(\Delta F)$ is
trivially zero. In order to determine the relevance or irrelevance of
the disorder we need to calculate:
\begin{equation}
{\rm slope}=\frac{d}{dx}\frac{\sigma(\Delta F)}{[\Delta F]}|_{x=0}.
\end{equation}
Here $x$ denotes the concentration of antiferromagnetic bonds.
At a standard fixed point we would expect the slope to follow a 
power-law behavior $L^\lambda$ and in the case of $\lambda <0$ we
qualify the perturbation as irrelevant and as relevant if $\lambda>0$.
The calculation of the derivative with respect to $x$ is performed by
introducing a {\it single} antiferromagnetic bond into the lattice.
For the case of the bimodal distribution that we consider it turns out
that there are only two non-equivalent positions on the torus. Hence,
we can evaluate both $[\Delta F]$ and $\sigma(\Delta F)$ exactly. A
single antiferromagentic bond corresponds to a given value of $x$ and
since $r$ at $x=0$ is trivially zero we can numerically determine the
derivative. One should note that the error introduced by
numerically
determining the derivative is extremely small due to the large lattice
sizes and the corresponding tiny value of $x$.

Our results for the slope at $T=T_c$ are shown in Fig.~\ref{fig:slope}
for lattice sizes of $L=16,24,32,40,48$. Surprisingly the slope
increases with the lattice size $L$ and one could be led to the
conclusion that in the presence of this frustrating disorder the
perturbation is marginally relevant instead of marginally irrelevant as
is known to be the case for unfrustrating disorder. However, since our
results are numerically exact we can verify that the functional form is 
{\it not} a simple power-law. Instead, we find that the only acceptable
form is:
\begin{equation}
A\log(L)^\alpha L^{-\beta},
\label{eq:rgflow}
\end{equation}
with $\alpha\simeq0.36,\ \beta\simeq-0.07$ and $A=2.94$. This is a
functional form quite well-known for the correlation
functions~\cite{ludwig,shankar} in such disordered systems and as
recently shown by Cardy~\cite{cardy} it is common for many
low-dimensional models. Considering the precision of our numerical
results we believe this form to be correct up to small errors in
the exponents. If this is indeed the case the `flow' will not reverse
until lattice sizes of approximately L=200 are reached, a system size
not accessible by our methods. We do not attach any importance to this
length scale since it depends on the form of the disorder and various
other microscopic parameters. In the inset in Fig.~\ref{fig:slope} we show our
numerical results along with the extrapolation out to lattice sizes
of L=10000.

\section{Conclusion}
Applying the well-know mapping of the partition function of the
Ising model onto a pfaffian we have succeeded in evaluating the
finite-size renormalization group flow around the pure fixed point
in an almost exact numerical manner. We have used a bimodal
distribution of the disorder which includes frustration for which
the domain wall free energy as well as its standard deviation can
be evaluated exactly.
Our results show that very large
lattice sizes have to be used before it becomes clear that disorder,
even frustrating, is irrelevant. It is possible that the functional
form for the finite-size renormalization group flow
Eq.~(\ref{eq:rgflow}) can be verified using analytical methods.

\acknowledgements
We would like to thank D. Fisher for suggesting to look at the mapping
onto pfaffians and M.~Gingras for many useful discussions. It is also
a pleasure to thank M.~Mahajan and collaborators for kindly sending
us their source code.

%Nishimori line:
%\begin{equation}
%v=2p-1,\ \ \ v=\tanh\left(J/kT\right), \ \ \ w=v^2.
%\end{equation}
%$w_c=0.596\pm 0.008$, $p_c=0.886\pm0.003$, $T_c/J=0.975\pm0.006$.

\end{document}